\documentclass[11pt]{gh2013}

\usepackage{mathptmx}       
\usepackage{helvet}         
\usepackage{courier}        
\usepackage{makeidx}         
\usepackage{graphicx}        
\usepackage{multicol}        

\def\la{\mathrel{\mathchoice {\vcenter{\offinterlineskip\halign{\hfil
$\displaystyle##$\hfil\cr<\cr\sim\cr}}}
{\vcenter{\offinterlineskip\halign{\hfil$\textstyle##$\hfil\cr
<\cr\sim\cr}}}
{\vcenter{\offinterlineskip\halign{\hfil$\scriptstyle##$\hfil\cr
<\cr\sim\cr}}}
{\vcenter{\offinterlineskip\halign{\hfil$\scriptscriptstyle##$\hfil\cr
<\cr\sim\cr}}}}}
\def\ga{\mathrel{\mathchoice {\vcenter{\offinterlineskip\halign{\hfil
$\displaystyle##$\hfil\cr>\cr\sim\cr}}}
{\vcenter{\offinterlineskip\halign{\hfil$\textstyle##$\hfil\cr
>\cr\sim\cr}}}
{\vcenter{\offinterlineskip\halign{\hfil$\scriptstyle##$\hfil\cr
>\cr\sim\cr}}}
{\vcenter{\offinterlineskip\halign{\hfil$\scriptscriptstyle##$\hfil\cr
>\cr\sim\cr}}}}}

\begin{document}

\title*{The starbursts in the Milky Way}
\titlerunning{Milky Way starbursts}
\author{Ignacio Negueruela$^{1}$}
\authorrunning{I. Negueruela} 
\institute{$^{1}$Departamento de F\'{i}sica, Ingenier\'{i}a de Sistemas y
  Teor\'{i}a de la Se\~{n}al, Universidad de Alicante, Apdo. 99, E03080
  Alicante, Spain, \email{ignacio.negueruela@ua.es}}
%
%
\maketitle


\vskip -3.5 cm  
\abstract{High-mass stars are major players in the chemical and dynamical evolution of galaxies, and young massive clusters are the natural laboratories to study their evolution and their impact on star formation processes. Only in recent years have we become aware of the existence of numerous massive ($M_{{\rm cl}}> 10^{4}\:M_{\odot}$) clusters in our Galaxy. Here I give a review, rather biased towards my own research interests, of the observational and theoretical efforts that have led to a description of their properties, and present an overview of the two (perhaps three) starburst regions known outside the Galactic Centre neighbourhood: the Scutum Complex, its putative counterpart on the far side of the Long Bar, and the starburst cluster Westerlund~1.
}

\section{Massive clusters in the Milky Way}
Most of our current knowledge of the processes leading to star and cluster formation (e.g., \cite{mckee07}) is based on observations of the Solar neighbourhood, where stars are mostly formed in small clusters \cite{lada03}. Conditions in these star-forming regions are generally different from those in the inner regions of the Milky Way, and believed to be very different from those in starburst regions \cite{kennicutt12}. Not so long ago, studies of high-mass stars in the Milky Way were restricted to clusters with typical masses $M_{{\rm cl}}\sim 10^{3}\:M_{\odot}$, where poor sampling was a major drawback (e.g., \cite{stothers85}). The realisation in 1998 that the Arches cluster \cite{cotera96} was a massive young cluster came as a surprise, and was taken as evidence for a distinct mode of star formation close to the Galactic Centre \cite{serabyn98}. At the time, it was thought that all open clusters in the Milky Way had masses below $10^{4}\:M_{\odot}$ (e.g., \cite{larsen99}).

In recent years, however, in-depth studies have shown that some clusters and associations that had been known for long are more massive than previously thought. Good examples include the double cluster in Perseus (h \& $\chi$ Per), with a combined mass $M_{{\rm tot}}\approx 2\times10^{4}\:M_{\odot}$ \cite{currie10} or the Carina Nebula complex, which includes the massive cluster Trumpler~14, with a mass $M_{{\rm cl}}\approx 10^{4}\:M_{\odot}$ \cite{ascenso07}. In these cases, rather massive, compact clusters are surrounded by diffuse OB associations. A slightly different spatial configuration is found in the compact association Cygnus~OB2, for which Wright et al.\ (\cite{wright10}) estimate a total mass $M_{{\rm tot}}\approx 3\times10^{4}\:M_{\odot}$. The association is very compact \cite{comeron08} and may present an amount of subclustering \cite{negueruela08}, but the stellar density is very low ($10^{2}\:M_{\odot}\,{\rm pc}^{-3}$; \cite{wright10}) compared to typical densities in a starburst cluster ($10^{5}\:M_{\odot}\,{\rm pc}^{-3}$). Cyg~OB2 seems to have formed with this low density \cite{wright14}. Perhaps massive star formation in the Outer Milky Way tends to happen in the form of dispersed associations; perhaps it is a hierarchical process that may result in different sorts of dynamical structures \cite{gieles11, kruijssen12}. An even more massive association is likely to result from the embedded star-forming region W49A, which is expected to produce several moderately-massive clusters \cite{homeier05alves}.

Very compact massive clusters also exist in the Milky Way. Perhaps the most well-studied is NGC~3603, which provides a good example of how our knowledge of compact, distant clusters has changed as technology developed. Its central object, HD~97950, was recognised as a compact group of stars about 40 years ago \cite{walborn73,moffat74}. In 1983, Moffat (\cite{moffat83}) identified several early O-type stars using photographic spectra. In 1994, using the Planetary Camera on board {\it HST}, Moffat et al. (\cite{moffat94}), resolved the central object into three luminous Wolf-Rayet stars and a large number of O-type stars. The Wolf-Rayet stars have later been found to be very massive stars in the H-core burning phase -- one of them a binary containing the most massive star with a dynamical mass determination (A1a with $M_{{*}}= 116\pm31\:M_{\odot}$; \cite{schnurr08}). Using adaptive optics at the VLT, Harayama et al. (\cite{harayama08}) derive a mass in the range $M_{{\rm cl}}= 1$\,--\,$1.6\times 10^{4}\:M_{\odot}$ with indications of a top-heavy initial mass function (IMF). However, the actual mass is heavily dependent on the distance to the cluster. Even though the reddening is not very high, $E(B-V)=1.39$ \cite{melena08}, there is strong evidence for a non-standard reddening law, leading to a large dispersion in estimated distances.

As we move to higher extinctions, this problem becomes dominant. The most extreme cases are the clusters close to the Galactic Centre. Leaving aside the Galactic Centre ``cluster'' (see, e.g., \cite{genzel10}), there are two very obscured compact, massive aggregates in this area, the Quintuplet \cite{hussman12} and Arches \cite{figer02} clusters. These clusters are affected by very heavy and variable reddening, to the point that it could give rise to colour terms even in the infrared $JHK$ system \cite{espinoza08}. Even worse, the shape of the extinction law becomes very difficult to determine, because the stars are completely invisible at optical wavelengths. Most recent studies conclude that the extinction law towards the Galactic Centre is very different from that generally used for Milky Way lines of sight with moderate reddenings (e.g., \cite{nishiyama09}). As a consequence of these problems, the total mass and IMF of the very young ($\tau\sim2$~Myr) and compact Arches clusters are very poorly determined. Stolte et al. (\cite{stolte05}) found evidence for a low-mass-truncated mass function that would imply a relatively small cluster mass $M_{{\rm cl}}\la10^{4}\:M_{\odot}$. Espinoza et al. (\cite{espinoza08}) argued for a much flatter mass function that implied a measured $M_{{\rm cl}}\approx 2\times10^{4}\:M_{\odot}$ in the central 0.4~pc and a stellar density $\approx 2\times 10^{5}\:M_{\odot}\,{\rm pc}^{-3}$. If the mass function is not truncated, this would imply a total cluster mass $M_{{\rm cl}}\ga 5\times10^{4}\:M_{\odot}$. More recently, Clarkson et al. (\cite{clarkson12}) used proper motions to separate actual members from the field population and found a mass $M_{{\rm cl}}\approx 1.5\times10^{4}\:M_{\odot}$ in the central 1.0~pc, giving support to a top-heavy present-day mass function.  The total mass implied would be $M_{{\rm cl}}\sim 3\times10^{4}\:M_{\odot}$.

All these examples show that clusters with masses  $M_{{\rm cl}}\approx 2\times10^{4}\:M_{\odot}$ are not unusual in the Milky Way. Moreover, the existence of a significant number of old open clusters with relatively high present-day masses suggests that they may be even more common than our current knowledge would suggest. For example, the intermediate-age open cluster M\,11 ($\tau=200$~Myr), with a present-day mass estimate of  $M_{{\rm cl}}\approx 1.1\times10^{4}\:M_{\odot}$ \cite{santos05}, is less than 2~kpc away from the Sun, suggesting that its progenitor is unlikely to be a rare occurrence. The heavily obscured cluster GLIMPSE-CO1 is likely a much more massive intermediate-age cluster (dynamical $M_{{\rm cl}}= 8\pm3\times10^{4}\:M_{\odot}$), though the possibility of a globular cluster crossing the Galactic disk has not been ruled out yet \cite{davies11}.

\section{The Scutum Complex}
\label{sec:scutum}

Since extinction is a major hindrance to locating Milky Way massive clusters, infrared surveys have been widely used to search for them (e.g., \cite{dutra03,chene12}), but O-type stars are not very bright in the near infrared. Red supergiants (RSGs), on the other hand, are extremely bright in the infrared, reaching $M_{K}\approx -9$ to $-12$~mag. Searches for large concentrations of RSGs led to the discovery of several clusters in a small region of the Galactic Plane, between $\ell=25^{\circ}$ and $30^{\circ}$.

The first cluster found, RSGC1 \cite{figer06}, is very heavily obscured ($A_{K_{{\rm S}}}>2$), and only visible in the infrared. With a dynamical distance estimation $d\approx6.5$~kpc, it should have an age $\tau=12\pm2$~Myr and a mass $M_{{\rm cl}}\sim 3\times10^{4}\:M_{\odot}$ \cite{davies08}. Though its main sequence may be barely identifiable in colour-magnitude diagrams \cite{froebrich13}, its mass has to be estimated by comparing the number of RSGs with the results of population synthesis models (see, e.g., \cite{clark09a}), and its parameters remain uncertain.

The cluster RSGC3 is less obscured, and its RSGs can be seen in the $R$ and $I$ bands. With a dynamical distance close to $d\approx6$~kpc ($v_{{\rm LSR}}\approx +95\:{\rm km}\,{\rm s}^{-1}$), it has an estimated mass $M_{{\rm cl}}\sim 3\times10^{4}\:M_{\odot}$ \cite{clark09b}. A systematic search for RSGs in its neighbourhood to determine its actual extent found a large number of RSGs with similar velocities, including at least another cluster \cite{negueruela11}. A third, more obscured cluster, Alicante~10, was found $\sim18^{\prime}$ South of RSGC3 \cite{gonzalez12}. More complete searches with the multi-object spectrograph AAOmega by our group (Dorda et al., in prep.) would place the number of RSGs within $30^{\prime}$ ($\sim 50$~pc at 6~kpc) of RSGC3 at $\ga70$, implying a total mass $\sim10^{5}\:M_{\odot}$ for the putative association. Interestingly, this agglomerate lies only $90^{\prime}$ away from W43, one of the most massive  giant star forming regions in the Milky Way \cite{nguyen11}, which has a parallax distance of $5.5^{+0.4}_{-0.3}$~kpc \cite{zhang14}. 

The open cluster Stephenson~2 \cite{stephenson} turned out to contain the highest number of RSGs, about 25 \cite{davies07}. Even though it lies behind a dust layer (LDN 515), it is less affected by extinction than the other RSG clusters. Observations of main-sequence stars indicate a reddening $E(J-K_{{\rm S}})=1.7$ \cite{gonzalez13,froebrich13}. The dynamical distance $d\approx6$~kpc ($v_{{\rm LSR}}\approx +109\:{\rm km}\,{\rm s}^{-1}$) \cite{davies07} is compatible with other distance estimates \cite{ortolani}, suggesting an age $\approx17\pm3$~Myr and a mass of at least $4\times10^{4}\:M_{\odot}$.

A comprehensive search for RSGs in the vicinity of Stephenson~2 led to the discovery of many supergiants with radial velocities similar to the cluster average \cite{negueruela12}. In view of this, the limits of Stephenson 2 cannot be clearly defined. A compact core, with radius $\la2.5^{\prime}$ and containing about 20 RSGs, is surrounded by an extended association
that merges into a general background with an important over-density
of RSGs. This over-density, which had already been noticed \cite{lopez}, extends preferentially to the North East of the cluster (i.e., along the Galactic Plane) for more than one degree. New observations reveal more than 100 RSGs in this over-density (Dorda et al., in prep.), indicative of an underlying mass of a few $\sim10^{5}\:M_{\odot}$ in young stars.

At present we cannot claim to understand the nature of the Scutum Complex. It is highly unlikely to represent a single star forming region, since it would then span more than 400~pc along the Galactic Plane. The cluster agglomerate surrounding RSGC3 seems rather compact and could easily represent a more massive version of the typical OB associations, such as Per~OB1 \cite{humphreys78}. The over-density in the vicinity of Stephenson~2 extends from $\ell\approx26^{\circ}$ to close to $28^{\circ}$. Its possible extension towards RSGC3 (at $\ell=29.2^{\circ}$) needs to be investigated. The location of the Complex, near the base of the Scutum Arm, has led to the hypothesis that it could represent a major starburst caused by the interaction of the tip of the Long Galactic Bar with the spiral arm \cite{lopez,davies07}. Though this hypothesis is very appealing, and the only compelling explanation proposed so far for the existence of the Complex, several complications must be considered:

\begin{itemize}
\item Firstly, there are no valid spiral tracers in this direction with LSR velocities significantly higher than that of Stephenson~2, introducing the possibility that the large over-density of RSGs simply represents a projection effect of objects with radial velocities close to the terminal velocity. However, this seems unlikely in view of the observed amounts of extinction \cite{negueruela12}.
\item  Secondly, kinematic distances assume no strong deviations from the Galactic rotation curve, while the perturbing effect of the Galactic Bar could easily lead to large departures. Understanding if this is the case would need independent distance determinations. 
\item Finally, other distance determinations must assume an extinction law
in the infrared. Though estimates of the distance to Stephenson~2 (the less obscured cluster) seem to agree with a standard extinction law, a careful determination of its shape would be desirable.
\end{itemize}

\section{The far side of the Bar}
\label{sec:far}

If the conglomerate of massive clusters towards the base of the Scutum Arm is related to the perturbing action of the Long Bar, a similar concentration should be seen towards its other tip. One of the main difficulties in the search for these clusters is the uncertainty in the angle formed by the Bar and thus in the position of its far end \cite{gonzalez12b}. Davies et al. (\cite{davies12}) found that the obscured cluster Mercer~81, located at $\ell\approx338^{\circ}$, is a young ($\tau\approx4$~Myr) and massive  $M_{{\rm cl}}> 10^{4}\:M_{\odot}$ cluster beyond more than 40~mag of extinction, at a distance $\sim11$~kpc, and claimed that it could signal the presence of a starburst region. 

Another likely massive young cluster is [DBS2003]~179, for which Borissova et al. (\cite{borissova12}) find a mass $M_{{\rm cl}}> 2\times 10^{4}\:M_{\odot}$ and a distance $\approx8$~kpc. This cluster is located much closer to the Galactic Centre, at $\ell\approx348^{\circ}$, but the amount of extinction is lower ($A_{V}\approx18$). Just South of the Galactic Plane in this direction, infrared maps reveal the existence of a very large gap in the dust distribution. At $\ell\approx349^{\circ}$, the open cluster VdBH~222 seems to lie in a low-extinction window. Our recent work shows that it contains 10 RSGs, and is a slightly less massive counterpart to the clusters at the base of the Scutum Arm (Marco et al., submitted). The photometric analysis favours the long kinematic distance of $\sim10$~kpc, implying an age $\sim16$~Myr and a mass approaching $M_{{\rm cl}}\approx 2\times 10^{4}\:M_{\odot}$. The relatively low extinction to the cluster suggests that it cannot be behind the Long Bar.

\section{Westerlund~1}
\label{sec:wd1}
The two possible starburst regions discussed above contain a number of clusters and an extended population. Contrarily, the most massive cluster known in the Milky Way seems to have been born in almost complete isolation. With a population of $>70$ supergiants of spectral types ranging from O to M \cite{negueruela10}, Westerlund~1 (Wd~1) is the prime laboratory for the study of high-mass star evolution. Its population of Wolf-Rayet stars \cite{crowther06} and high-mass interacting binaries \cite{ritchie09} can provide stringent tests on theoretical models. The spread in brightness of OB supergiants is consistent with a single age \cite{negueruela10}, in agreement with the analysis of the main sequence width that found an age spread $<0.4$~Myr \cite{kudryavtseva12}.  

A direct extrapolation of the number of massive stars detected would suggest a mass $M_{{\rm cl}}\approx 10^{5}\:M_{\odot}$ for a standard IMF \cite{clark05}. The radial velocity dispersion of 10 bright stars resulted in a mass  $M_{{\rm cl}}\approx 1.5\times 10^{5}\:M_{\odot}$ \cite{mengel09}. Many of the stars used, however, are radial velocity variables \cite{clark10}, and this value must be an overestimate. Direct star counts in the infrared give a lower limit of  $5\times 10^{4}\:M_{\odot}$ \cite{gennaro11}. This mass estimate, however, relies on an assumed age and distance. There seems to be some disagreement between the age and distance estimate obtained from the post-main-sequence population and those obtained from the pre-main sequence, a situation found in many young clusters \cite{negueruela10}. 

Again, these parameters are also dependent on the extinction law adopted. Using multi-epoch high-resolution spectroscopy of a large sample of cluster members, Clark et al.\ (submitted) are able to derive an average cluster radial velocity, which leads to a surprisingly low kinematic distance, inconsistent with values $\ga5$~kpc obtained from the high-mass stellar population \cite{crowther06, negueruela10}. A rather unusual extinction law would be necessary to bring the cluster to this low distance.

\section{Conclusions}

In the past fifteen years, we have discovered almost fifteen open clusters more massive than $10^{4}\:M_{\odot}$ in the Milky Way. This has changed our perception of the Galactic cluster system. The concentration of these massive clusters towards the Sun suggests that many more are waiting to be discovered, in most cases hidden behind huge amounts of extinction.

Most of these clusters are relatively isolated, or surrounded by diffuse associations. Apart from the massive population near the Galactic Centre, the strongest concentration is the Scutum Complex, which extends over $\sim400$~pc. At present, it is unclear if it represents a single entity, but the associations found surrounding Stephenson~2 and RSGC3 are likely the most massive ones known in the Milky Way, even if they are not connected. The most likely explanation for the existence of these large star-forming regions is triggering by the tip of the Long Bar. If so, a similar complex should exist on the other side of the Bar. Several massive clusters have been found in that direction. At least one of them, VdBH~222 is a cluster of red supergiants surrounded by a diffuse association, similarly to its near-side counterparts.

Even though Stephenson~2 is very massive, Westerlund~1 is likely to be the most massive cluster in the Milky Way. At an age between 4 and 6~Myr, its unique evolved population makes it the ideal laboratory to study high-mass star evolution. Wd~1 seems to have formed in isolation, and its radial velocity would place it at a distance difficult to reconcile with its observed population. These characteristics might be hinting at some violent dynamical process as the reason for its formation.

\begin{acknowledgement} 
I would like to thank my collaborators in cluster research, Drs. Simon Clark, Carlos Gonz\'alez-Fern\'andez and Amparo Marco for many years of fruitful collaboration and comments on the manuscript. This research is partially supported by the Spanish Ministerio de Econom\'{\i}a y Competitividad under grants AYA2010-21697-C05-05 and AYA2012-39364-C02-02.
\end{acknowledgement}

\end{document}